\documentclass[twocolumn,amsmath,amssymb,floatfix,prb,preprintnumbers,footinbib]{revtex4-1}
\newlength{\stdwidth}
\usepackage[dvips]{graphicx}

\usepackage{bbm}
\usepackage{amsfonts}
\usepackage{amsmath}
\usepackage{graphicx}
\usepackage[utf8]{inputenc}

\usepackage{hyperref}

\newcommand{\subref}[2]{\ref{#1}({#2})}
\newcommand{\sub}[1]{{#1}}

\usepackage[usenames]{color}
\definecolor{contentlistcolor}{rgb}{.5,.5,.7}

\def\mathvecfont#1{\textbf{\em #1}}
\newcommand{\myvec}[1]{\mathvecfont{#1}}

\newcommand{\ci}{\mathrm{i}}
\newcommand{\e}{\mathrm{e}}

\newcommand{\UP}{\uparrow}
\newcommand{\DOWN}{\downarrow}
\newcommand{\RM}{+}
\newcommand{\LM}{-}
\newcommand{\RMLM}{\pm}

\newcommand{\textfrac}[2]{{#1}/{#2}}

\newcommand{\la}{\lambda}

\begin{document}
 \title{
Switching spin and charge between edge states in topological insulator constrictions }

\author{Viktor Krueckl}
\affiliation{Institut f\"ur Theoretische Physik, Universit\"at Regensburg, D-93040 Regensburg, Germany}

\author{Klaus Richter}
\affiliation{Institut f\"ur Theoretische Physik, Universit\"at Regensburg, D-93040 Regensburg, Germany}

\date{\today}

\begin{abstract}
Since the prediction of a new topological state of matter in graphene,
materials acting as topological
insulators have attracted wide attention.
Shortly after the theoretical proposal for a mercury telluride (HgTe)-based  
two-dimensional topological insulator,
the observation of the quantum spin Hall effect and 
non-local edge transport brought
compelling experimental evidence for quantized conductance due to edges states.
The spin orientation and propagation direction of such helical edge states
are inherently connected, providing protection against backscattering.
However, these features conversely render controlled spin operations such as
spin switching difficult.
Here we therefore propose constrictions as connectors between opposite edge (and spin) states in HgTe. 
We demonstrate how the coupling between edge states, which overlap in
the constriction, can be employed both for steering the charge flow into different
edge modes and for controlled spin switching. This gives rise to a three-state charge and spin
transistor function.
Unlike in a conventional spin transistor, the switching does not rely on 
a tunable Rashba spin-orbit interaction, but on the energy dependence 
of the edge state wavefunctions.
Based on this mechanism, and supported by numerical transport calculations,
 we present two different ways to control 
spin- and charge-currents depending on the gating of the constriction. 
\end{abstract}
\maketitle

%
%
%%%%%%%%%%%%%
\begin{figure}[b]
\centering
\includegraphics[width=\stdwidth]{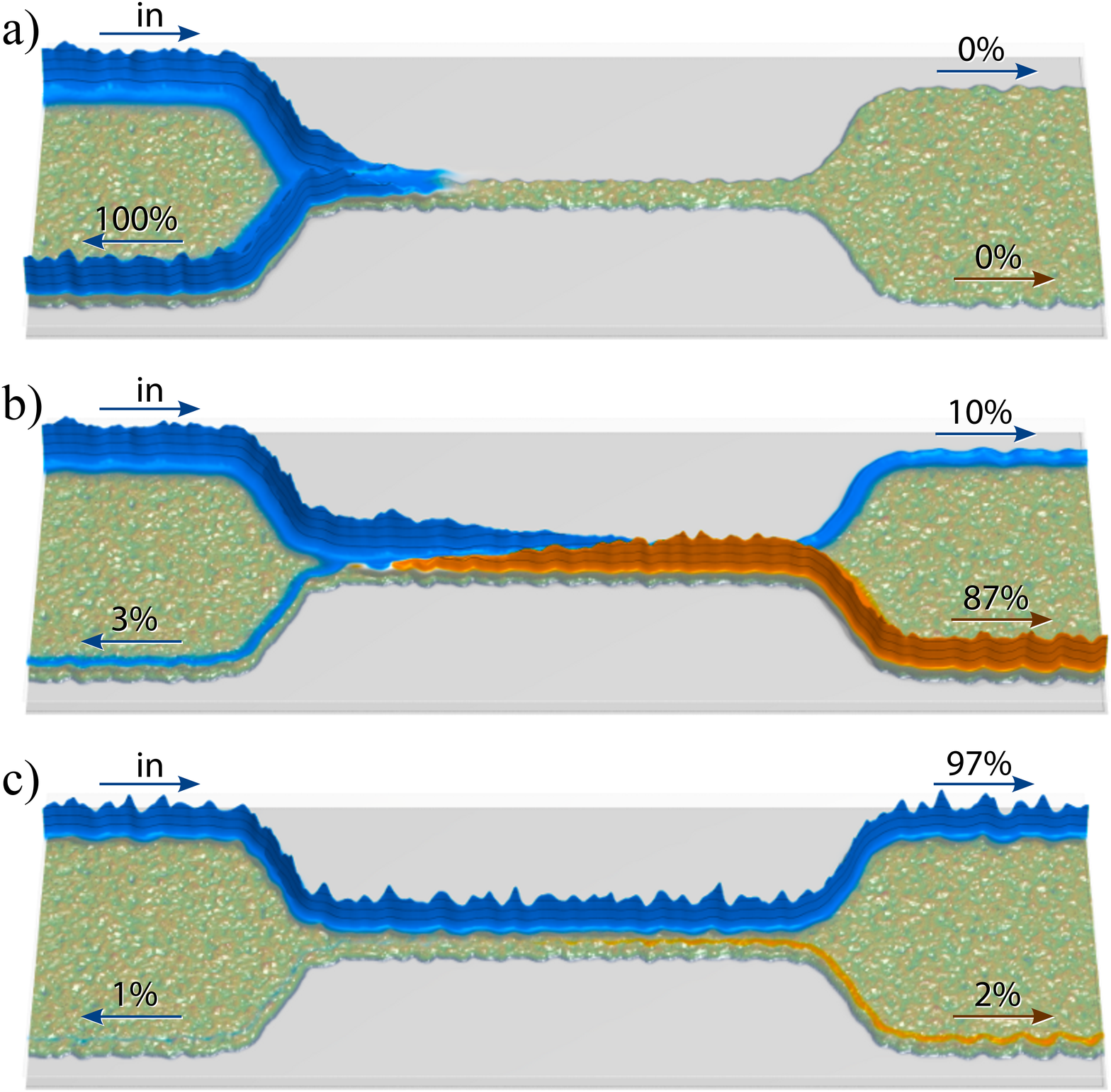}
\centering
\includegraphics[width=\stdwidth]{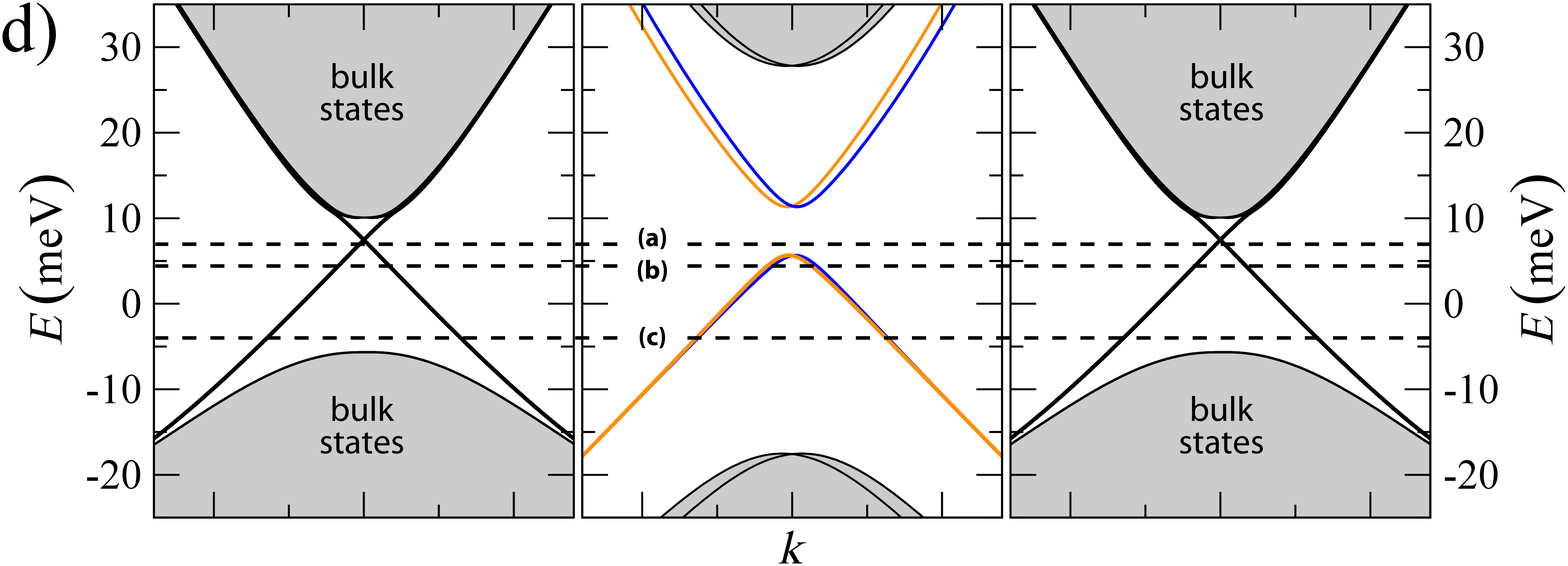}
\caption{
({\em color online})
{{ Local density of states and band structure of a HgTe constriction for different switching conditions}} --
\sub{a-c}, Spin-resolved local density of states for charge carriers entering a disordered constriction from the upper left edge. 
The colour code indicates the spin polarization of the state, with blue representing the upper subblock ($\UP$) and orange the
lower subblock ($\DOWN$).
\sub{a}, The chemical potential in the narrow region is chosen within the confinement
induced gap ($\mu = 7\;{\rm meV}$), leading to a perfect reflection of the incoming state.
\sub{b}, For an energy closely below the gap ($\mu = 4\;{\rm meV}$) the effective spin-orbit coupling
is pronounced enough to produce a precession from spin-up to spin-down and the associated switching
from the upper (left) to the lower (right) edge.
\sub{c}, At low energies, distinctly below the gap ($\mu = -5\;{\rm meV}$), the edge states are predominantly localized at the respective
boundary in the constriction, leading to a reduced effective
SOI, and hence the incoming spin-up states leave the constriction unrotated. 
\sub{d}, Sketch of the HgTe band structure for the sequence of a wide (left panel, $W = 1000\;{\rm nm}$), 
 narrow (middle panel, $W = 100\;{\rm nm}$) and wide (right) lateral confinement.
 Dashed lines mark the energies of the local density of states in panels \sub{a-c}.
}
\label{figdos}
\end{figure}
%%%%%%%%%%%%%
%
%

Realizations of topological insulators\cite{Kane2005b,Kane2005a} with their exceptional electronic and transport
properties exist both in three dimensions, e.g. the surface states of BiSe\cite{Fu2007}
or Bi$_2$Te$_3$\cite{Qu2010, Xiu2011}, and in
two dimensions with HgTe as prominent example.
The topological properties of HgTe/CdTe heterostructures  are attributed to the 
combination of an inverted band structure of the HgTe layer (the light-hole band as 
conduction band and the heavy-hole as the topmost valence band) and the regular band 
structure of CdTe.
Quantized transport along the HgTe boundaries can be conveniently explained by an
edge channel picture:\cite{Buttiker1988}
Two states with opposite spin orientation propagate along opposite device edges 
in the same direction and thus lead to a quantized conductance of $2 e^2/h$.
Due to the spatial separation of the spin-states the spin-orbit coupling is suspended,
and the system geometry can be employed for spin selection\cite{Roth2009}.

Spin-selectivity is also a crucial element of the Datta-Das spin transistor 
proposal\cite{Datta1989}, where charge flow is controlled electrically through the gate-dependent 
Rashba spin orbit interaction\cite{Bychkov1984} (SOI) in a conventional two-dimensional 
semiconductor heterostructure placed in between ferromagnetic contacts.
Its realization, however, turns out to be difficult owing to spin relaxation in the 
semiconductor heterostructure and interfacial effects such as
the conductivity mismatch between the different materials\cite{Schmidt2000}. 
HgTe-based topological insulators appear to be promising candidates for spin processing
devices since they also can be gated and exhibit considerable SOI but, on the contrary,
are composed of a single material class only. Moreover, the one-dimensional (1d) nature of 
their edge states suppresses orbital effects present in bulk conductors,
leading to high spin polarizations and to a much better (spin) switching quality.

To our knowledge there have been only a few proposals for spin-transistors 
based on two-dimensio\-nal topological insulators.
Two of them rely on spin switching with a magnetic field at a
pn-junction\cite{Akhmerov2009} or in an Aharonov-Bohm interferometer\cite{Maciejko2010}.
Recently it has further been suggested, within a phenomenological model, that separate gating 
of the two branches of an Aharonov-Bohm interferometer allows for manipulating 
charge- and spin-transport\cite{Dolcini2010}.
By contrast, our present proposal relies on an electrical operation by gates
on a single HgTe nano-constriction,
which up to now was only considered for charge current switching\cite{Zhang2011}.

In this manuscript we demonstrate how topological edge states can be selectively switched 
in an elongated constriction etched out of a HgTe heterostructure, leading to an
integrated three-state charge- and spin-transistor of high fidelity.
An incoming spin-polarized (upper edge) state can either be reflected back to the lower edge,
as shown in Fig. \subref{figdos}{a}, or transmitted through the narrow part.
Back-reflection into the opposite spin channel at the same edge is forbidden,
as the absence of a magnetic field implies time-reversal symmetry~\cite{Hasan2010}.
Within the constriction the SOI between the edge channels is reactivated
due to a finite overlap between right moving edge-states on upper and lower side, giving rise to spin precession
that is tunable by a gate.
This allows for steering the spin orientation of the electrons which leave the constriction,
and thereby their further path:
An incoming spin-up state will leave the system either by swapping the edge
(with a simultaneous spin-flip), as shown in Fig. \subref{figdos}{b},
or by remaining in its spin and edge state as shown in Fig. \subref{figdos}{c}.

%%%%%%%%%%%%%%%%%%%%%%%%%%%%%%%%%%%%%%%%%%%%

In the following, we first introduce a 1d model
which explains all relevant spin and charge transport features arising from the 
constriction.
Subsequently, we check the predictions of this model by numerical
quantum transport calculations which include random impurity potentials and
rough confining potentials to account for material inhomogeneities and
etching imperfections. 
Furthermore, we analyse the peculiar switching properties by means of a top-gate and side-gates
acting on the constriction.

We describe the electronic properties of the underlying HgTe heterostructure by the four band
Hamiltonian\cite{Bernevig2006a, Roth2009}
%
%
%%%%%%%%%%%%%%%
\begin{equation}
H = {\small \left ( \begin{smallmatrix}
M-(B+D) \myvec k^2 & A k_+ & - \ci R k_- & -\Delta \\
A k_- & -M+(B-D) \myvec k^2 & \Delta & 0 \\
\ci R k_+ & \Delta& M-(B+D) \myvec k^2  & -A k_-\\
-\Delta& 0 & -A k_+& -M+(B-D) \myvec k^2 \\
\end{smallmatrix}
\right ) }
\label{Hhgte}
\end{equation}
%%%%%%%%%%%%%%%
%
%
where $k_\pm = k_x \pm \ci k_y$ and $\myvec k^2 = k_x^2 + k_y^2$.
This Hamiltonian contains the commonly used time-inverted $2\times2$ blocks
for the composite states of the heavy-hole and electron bands\cite{Bernevig2006a}.
Additionally, we take into account the leading order SOI terms $\Delta$ and $R$,
due to bulk-inversion asymmetry (BIA)
and structure-inversion asymmetry (SIA)\cite{Rothe2010}, respectively.
(See the supplementary material for the values used for $\Delta$, $R$ and the further material
parameters $A$, $B$, $D$ and $M$.)

%%%%%%%%%%%%%%%%%%%%%%%%%%%%%%%%%%%%%%%%%%

Based on the Hamiltonian (\ref{Hhgte}) we first derive an effective Hamiltonian for an infinite strip
of constant width $W$.
We chose the lead to point in $x$-direction and search
for the transversal eigenfunctions $\psi(y)$
which separately fulfill the boundary conditions  $\psi(y \le 0) = 0$
at the lower side and $\psi(y\ge W) = 0$ at the upper side.
For a very wide confinement one can neglect the influence
of the opposite boundary on the edge states, leading to a full spin-polarization.
Then the resulting states can be classified by their subblock into up $(\UP)$ and down $(\DOWN)$,
as well as their propagation direction into right movers $(\RM)$ and left movers $(\LM)$.
Accordingly, the right moving state of the upper subblock is given by
%
%
%%%%%%%%%%%%%%%
\begin{equation}
\psi_{\UP}^{\RM}(y) \propto 
\left (
 \e^{\la_1^{\RM} (y - W)} -  \e^{\la_2^{\RM} (y - W)}
\right ) 
\begin{pmatrix}
1, & -\xi^{\RM}, & 0, & 0
\end{pmatrix}
\end{equation}
%%%%%%%%%%%%%%%
%
%
and the right moving state of the lower subblock by
%
%
%%%%%%%%%%%%%%%
\begin{equation}
\psi_{\DOWN}^{\RM}(y) \propto 
\left (
 \e^{-\la_1^{\RM} y} -  \e^{-\la_2^{\RM} y}
\right ) 
\begin{pmatrix}
0, & 0, & 1, & \xi^{\RM}
\end{pmatrix},
\end{equation}
%%%%%%%%%%%%%%%
%
%
with two different decay exponents $\la_1^{\RM},\la_2^{\RM} > 0$ and
$\xi^\RM$ the weight of the second spinor entry in the respective subblock.
Then the left moving states are obtained by complex conjugation and
by substituting $y \rightarrow W - y$ and $k_x \rightarrow - k_x$
which corresponds to inter-changing $\la_{1/2}^{\RM} \rightarrow \la_{1/2}^{\LM}$ and
$\xi^{\RM} \rightarrow \xi^{\LM}$.
(The functions are described in detail in the supplementary material.)
From these properties we can derive an effective 
1d Dirac Hamiltonian 
$H_{\mathrm{eff}\UP/\DOWN} = c \mp a \sigma_x k_x$
for a single spin subblock
with a velocity $a\;=\;A\;\sqrt{\textfrac{(B^2 - D^2)}{B^2}}$
and an energy offset $c\;=\;-\textfrac{D M}{B}$.
For a wide strip this is in perfect agreement with the
full band structure in the vicinity of the band crossing shown in the left and right panel of
Fig.~\subref{figdos}{d} for $W = 1000~\mathrm{nm}$.

For decreasing width $W$, the edge states at opposite boundaries start to overlap,
leading to a mass like gap in the 1d Hamiltonian.
By invoking simultaneously the boundary conditions for the upper and lower side  
the size of the effective mass gap is given by\cite{Zhou08}
\begin{equation}
m \approx \frac{2 | A (B^2 - D^2) M | }{ B^3 ( A^2  B - 4 (B^2 - D^2) M) } \e^{- \la_1^\RM W}.
\label{mass}
\end{equation}
Additionally, the suppression of the SOI for the edge states is suspended for
small $W$.
Neglecting the rapidly decaying terms proportional to  $\e^{\la_2^\RMLM y}$ in the wavefunction, the
overlap due to BIA can be stated as
\begin{equation}
\delta^{\RMLM} \approx -\frac{4 \e^{-\la^\RMLM W} \la^\RMLM W \xi^\RMLM}{1 + (\xi^\RMLM)^2} \Delta.
\label{effSOI}
\end{equation}
The effect of SIA on the edge states within the band gap is negligible small.
Combining the effective mass $m$ of Eq. (\ref{mass}) and the effective SOI $\delta^{\RMLM}$
of Eq. (\ref{effSOI}), we can compose a 1d effective Hamiltonian
%
%
%%%%%%%%%%%%%%%%%
\begin{equation}
H_{\rm eff} = 
\begin{pmatrix}
c + m & - a k_y & \delta_{\rm m} & \delta_{\rm p} \\
- a k_y & c - m & -\delta_{\rm p} & -\delta_{\rm m} \\
\delta_{\rm m}  & -\delta_{\rm p} & c + m & a k_y \\
\delta_{\rm p}  & -\delta_{\rm m}  & a k_y & c -m
\end{pmatrix}
\label{Heff}
\end{equation}
%%%%%%%%%%%%%%%%%
%
%
with $\delta_{\rm p/m} = (\delta^\RM \pm \delta^\LM)/2$.
The band structure of $H_{\rm eff}$ exhibits a mass gap and a variable,
energy dependent effective spin-orbit splitting which is strongest close to the avoided band crossing 
shown in the middle panel of Fig.~\subref{figdos}{d} for a constriction of width $W = 100~\mathrm{nm}$.
Note that the SOI of the model and the corresponding splitting in the band structure is
slightly overestimated compared to the result of the full Hamiltonian, due to the restriction
to only four basis functions.

%
%
%%%%%%%%%%%%%
\begin{figure}[tbhp]
\centering
\includegraphics[width=\stdwidth]{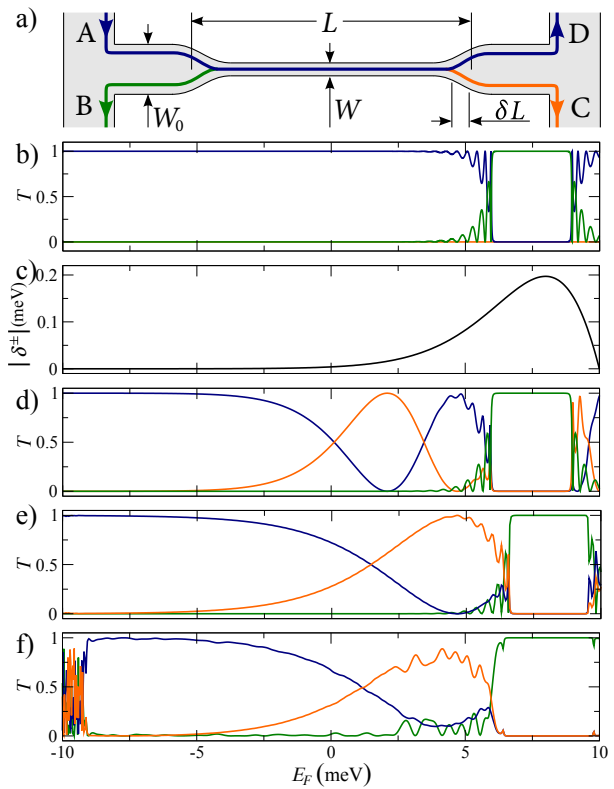}
\caption{
({\em color online})
{{Quantum transmission through a constriction: one-dimensional model vs. full quantum mechanics}} --
\sub{a}, Sketch of the constriction geometry. The coloured paths illustrate the differently scattered edge channels entering through lead A
 and leaving the system at lead B, C or D.
\sub{b}, Transmission through a constriction neglecting SOI calculated for the 1d model Hamiltonian, see text.
Different colours mark transmission paths according to colour coding in (\sub{a}).
\sub{c}, Energy dependence of the effective SOI, Eq. (\ref{effSOI}).
\sub{d}, Transmission through a constriction with SOI calculated by means of the 1d
model Hamiltonian (\ref{Heff}).
\sub{e, f}, Transmission obtained by a full numerical wave-packet calculation for (\sub{e}) a perfect constriction
and (\sub{f}) a realistic constriction 
with random impurity potential ($U_0 = 2\mathrm{meV}$) and rough walls ($W_r=20\mathrm{nm}$), the
same system as used in the calculation shown in Fig.~\subref{figdos}{a-c}.
Geometric properties of the constriction: width $W=100~\mathrm{nm}$, length $L=1900~\mathrm{nm}$ (${\delta}L = 50~\mathrm{nm}$),
width of bulk parts $W_0=1000~\mathrm{nm}$.
}
\label{figtransmission}
\end{figure}
%%%%%%%%%%%%%
%
%

In the following, we analyze within this model the transport properties for a constriction interconnecting two
bulk-like regions in an H-shaped HgTe heterostructure, as
depicted in Fig.~\subref{figtransmission}{a}.
Within the bulk bandgap the transport is exclusively carried by edge states.
Accordingly, there are three different paths 
entering at lead A and leaving the system at lead B (green), C (orange) or D (blue).

The 1d model for this setting equals a Dirac equation with a
position-dependent mass potential, if we neglect SOI.
For an abrupt change in width, this model can be solved analytically~\cite{Gomes2008}
leading to a pronounced drop of the transmission from A to D in the energy region of the confinement induced gap
 as well as strong Fabry-Pérot like transmission resonances for energies below the  gap.
In our 1d model calculations we use a more realistic smooth variation of the width given by two Fermi functions,
%
%
%%%%%%%%%%%%%%%%%%
\begin{equation}
W(x) = W_0 - \left ( 
\frac{W_0 - W}{1+\e^{(x-L/2)/{\delta}L}} - 
\frac{W_0 - W}{1+\e^{(x+L/2)/{\delta}L}}
\right ) \, .
\end{equation}
%%%%%%%%%%%%%%%%%%
%
%
Here $W$ is the width of the constriction of length $L$, $W_0$ is the width outside the constriction,
which is chosen wide enough to ensure a gapless Dirac spectrum,
and ${\delta}L$ denotes the length of the transition from $W_0$ to $W$, see Fig.~\subref{figtransmission}{a}.
The transmission through the smooth constriction, as shown in Fig.~\subref{figtransmission}{b},
is calculated using a discretized transfer-matrix approach.
As a result the Fabry-Pérot type resonances are strongly reduced while 
the transmission gap remains unaffected.
This simple model can readily be used to describe switching~\cite{Zhang2011}
of the output current between lead B or D upon tuning the Fermi energy, 
as shown in Fig.~\subref{figtransmission}{b}.

However, neglecting the SOI 
forecloses the possibility of a unique spin-flip mechanism as shown in the following.
With SOI the right moving spin-up and spin-down states hybridize 
within the constriction to a symmetric and an antisymmetric composite state
with a difference in the wave vectors of 
${\Delta}k \approx \vert \delta^\RM \vert / a$.
Since the overlap strongly depends on the extent of the edge-states, which again is mainly
determined by the decay exponents $\lambda_1^\RMLM$, 
the effective spin-splitting energy $\vert \delta^\RMLM \vert$ [in Eq. (\ref{effSOI})],  has a pronounced energy dependence
as shown in Fig.~\subref{figtransmission}{c}.
For low energies (larger $\lambda_1^\RMLM$) both states are very closely bound to opposite edges. Thus 
the effective SOI is very weak and the states preserve their
initial spin while traveling through the constriction [see also Fig.~\subref{figdos}{c}],
leading to a perfect transmission from lead A to D
as shown by the blue curve in Fig.~\subref{figtransmission}{d}.
At higher energies, close to the confinement induced bandgap,
the spin splitting is pronounced and already visible in the
band structure shown in the center panel of Fig.~\subref{figdos}{d}.
Hence the states undergo a spin precession when traversing the constriction.
Upon tuning the energy, i.e. $\vert \delta^\RMLM(E) \vert$ [see Fig.~\subref{figdos}{c}],
the precession frequency can be steered.
This spin rotation leads to a finite transmission from
lead A to C, shown as the orange line in Fig.~\subref{figtransmission}{d}.
For a given constriction length there is an energy (around $2~{\rm meV}$) where the incoming
spin is flipped leading to nearly perfect transmission into lead C. 
Note that the three different transmission probabilities displayed in Fig.~\subref{figtransmission}{d}
 show well separated and pronounced maxima up to unity for the various paths.
As a result the 1d model suggests that a HgTe constriction is perfectly suited to function as a three state
spin-orbit transistor, switching with excellent on-off ratio between the outgoing leads B, C and D.
At the same time, this system allows for controlled spin swapping, i.e. when choosing path A to C.

%%%%%%%%%%%%%%%%%%%%%%%%%%%%%%%%%%%

So far we employed a 1d model to demonstrate the various charge and spin
switching properties of a HgTe-based nano-constriction.
In the following we study the robustness of these effects for a realistic setting.
To this end we calculate the conductance governed by the four band Hamiltonian (\ref{Hhgte})
through constrictions in two-dimensions with additional random impurity potentials and rough walls.
We have extended an efficient numerical method to calculate electronic transport by means of
the time-evolution of wave-packets\cite{kramer2010} to arbitrary spin-orbit coupled systems.
The propagation is calculated by means of an expansion of the
time-evolution operator in Chebychev polynomials\cite{Tal-Ezer1984}, ensuring negligible numerical errors
for long propagation times\cite{krueckl2009}.
A detailed description of this algorithm and the considered models for impurity potential and edge roughness
can be found in the supporting material.

To check the validity of the 1d model we first apply the wave-packet transport algorithm to a clean system 
with the same parameters.
The resulting energy dependent transmissions into leads B, C and D are shown in Fig.~\subref{figtransmission}{e}.
As mentioned before, the 1d model Hamiltonian [Fig.~\subref{figtransmission}{d}] overestimates
the effective SOI.
Consequently the spin-precession is reduced in the full calculation, leading to maximally
a bit more than half a spin precession for the chosen geometry.
Note that, as in the 1d model, there is a distinct energy at which the spin 
precession produces an exact spin-flip, leading to a perfect transmission
from lead A to lead C [shown as the orange line in Fig.~\subref{figtransmission}{e}].
Also the transmission from lead A to B (green line) in the energy region of the confinement induced mass gap
and the transmission to lead D (blue line) is in good correspondence with the model.

In order to approach realistic, experimentally available conditions,
we add an impurity potential with $U_0=2\mathrm{meV}$ and a wall roughness of $W_r=20~\mathrm{nm}$
and calculate the transport through such a system.
The resulting energy dependent transmissions are summarized in Fig.~\subref{figtransmission}{f}.
Due to impurity induced energy variations within the constriction and fluctuations in its width, the
mass gap is enhanced as visible in the transmission from lead A to B (green line).
Most notably, the strength of the spin-flip mechanism is maintained, although the efficiency of the spin
flip process is slightly reduced by the impurity potential and the wall roughness:
The spin flip transmission [from lead A to C, orange line in Fig.~\subref{figtransmission}{f}]
no longer reaches $T=1$ [see also Fig.~\subref{figdos}{b}].
Nevertheless, this reduction only amounts to 20\% for a very strong perturbation as used in these
calculations.
Consequently we conclude that the switching properties of such a device are robust against
electrostatic impurities and persist in non perfectly etched heterostructures.

%
%
%%%%%%%%%%%%%
\begin{figure*}[tbhp]
\centering
\includegraphics[width=2\stdwidth]{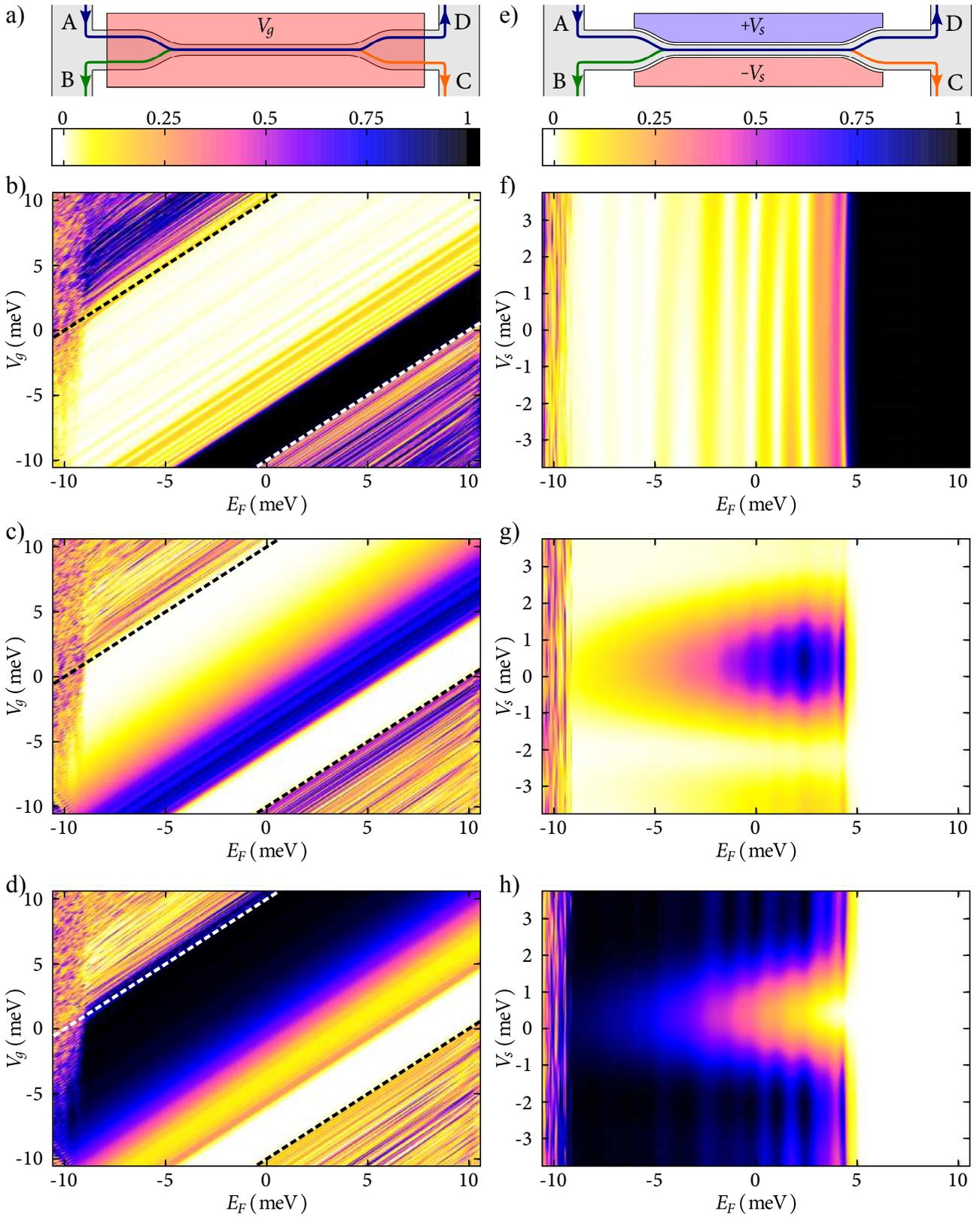}
\caption{
({\em color online})
{{ Top-gate- and side-gate-switching of a HgTe constriction}} --
\sub{a}, Scheme of the constriction (width $W=100~{\rm nm}$, length $L=1900~{\rm nm}$)
coupled to bulk parts (width $W_0=1000~{\rm nm}$) and subject to a top-gate (light red area).
For the transport calculation a random potential of $U_0=2~{\rm meV}$ and a wall roughness of 
$W_r=20~{\rm nm}$ are additionally included.
The impurity potential and wall roughness are visualized in Figs.~\subref{figdos}{a-c}.
Panels \sub{b-d} show the (colour coded) quantum transmission from lead A to the lead
B (\sub{b}), C (\sub{c}) and D (\sub{d}) as a function of the Fermi energy $E_F$ and the top-gate
voltage $V_g$ for 
the constriction as depicted in panel \sub{a}.
\sub{e}, Scheme of a smaller constriction (width $W=60~{\rm nm}$, wall roughness $W_r=10~{\rm nm}$, length $L=900~{\rm nm}$
coupled to bulk leads of $W_0=1000~{\rm nm}$) with two side-gates.
\sub{e-f}, Transmission from lead A to lead B (\sub{f}), C (\sub{g}) and D (\sub{h}) as a function
of $E_F$ and side gate voltage $V_s$.
Panels (\sub{b-d}) and (\sub{f-h}) feature a spin-transistor switching.
}
\label{figpotmap}
\end{figure*}
%%%%%%%%%%%%%
%
%

%%%%%%%%%%%%%%%%%%%%%%%%%%%%%%%%%%%%%%%%%%%%%%%%55

In the following we consider the possibility of controlled switching between edge currents by an
additional gate.
We model local gating that has been proven experimentally feasible~\cite{Roth2009},
by a position dependent potential which is switched on
outside of the confined region and spans a length of $3000~{\rm nm}$
[including the constriction, see Fig.~\subref{figpotmap}{a}].
Furthermore, a random impurity potential as well as edge roughness are again considered,
as specified above.
By means of our wave-packet algorithm we calculate the quantum transport through the device
as a function of both Fermi energy and gate voltage.
The resulting transmissions into leads $B$ to $D$ are shown in Fig.~\subref{figpotmap}{b-d}, where large (small)
transmissions are depicted by dark (bright) colours.
Note that all plots show universal conductance fluctuations for energies outside the bulk bandgap
[marked by dashed lines in Fig.~\subref{figpotmap}{b-d} and given approximately by
$E_F - V_g < -\vert M \vert$ and $E_F + V_g >  \vert M \vert $].
Within the bulk bandgap the device exhibits the switching properties.
The transmission from lead A to lead B, which is close to $1$ in the confinement induced
mass gap, is shown in Fig.~\subref{figpotmap}{b};
the transmission into lead C, accompanied by a spin-flip, in Fig.~\subref{figpotmap}{c}; and the transmission
to lead D without spin-flip in Fig.~\subref{figpotmap}{d}.
The Fermi energy within the whole system is pinned to a certain value and
can be globally tuned e.g.\ by means of another back-gate.
Assuming for example $E_F = 0$, the transmission can be steered between the
three different leads by changing the gate voltage locally on top of the constriction
and thereby changing the effective SOI.
In this case a perfect transmission to lead B is achieved for  $V_g = -7{\rm meV}$.
For $V_g = -4{\rm meV}$ the transmission to lead C is maximal, whereas for
$V_g = 5{\rm meV}$ the transmission to lead D approaches unity.
The fact that the spin flip process is slightly reduced [see Fig.~\subref{figpotmap}{c}] mainly stems from an average
potential difference between the upper and the lower wall, which
can be attributed to the electrostatic impurity potential.
This  leads to a deviation between the momenta of the right moving states from the two
different subblocks, and an associated reduction of the transition (and spin-flip) probability.

A similar effect occurs when we use two side-gates close to the constriction with an opposite 
applied voltage $V_s$, as sketched in Fig.~\subref{figpotmap}{e}.
This leads to different chemical potentials for the two spin channels close to the
upper and lower side of the constriction.
The results of a corresponding conductance calculation are summarized in Fig.~\subref{figpotmap}{f-h}.
Due to the different gating, the mass gap now shows up as a vertical stripe ($E_F > 5~{\rm meV}$), and
the constriction is isolating, independent of the gate voltage $V_s$,
as shown in Fig.~\subref{figpotmap}{f}.
If the constriction is tuned into the state conducting charge from left to the right ($E_F < 5~{\rm meV}$),
it works as a spin transistor controlled through the side-gate voltage $V_s$:
For low $\vert V_s \vert$ the states entering the device undergo a spin-flip within the constriction
whereas for larger $\vert V_s \vert$ the momenta of different spin states differ sufficiently to suppress
the spin precession.
For an illustration, compare the transmission from lead A to C in panel~(\sub{g})
with the transmission to lead D in panel~(\sub{h}).
As before, the calculations are performed under realistic conditions, including a random impurity potential and wall roughness
which reduce the spin flip probability.
Nevertheless, the remaining signal is still pronounced enough
to be observed experimentally in a single configuration.
The calculations (\sub{f-h}) were performed for a narrower
constriction ($W=60~{\rm nm}$) and the same amount of disorder
to demonstrate that the switching functionality is robust against down
scaling to a regime of a few 10 nanometres.
In view of Eq.~(\ref{effSOI}), the effective SOI increases with
decreasing width, allowing for faster spin precession and shorter constrictions
[$900~{\rm nm}$ in Fig.~\subref{figpotmap}{e-h}].

%%%%%%%%%%%%%%%%%%%%%%%%%%%%%%%%%%%%%%%%%

We have shown that a constriction joining together edge channels of the topological 
insulator HgTe acts as a transistor with unique charge and spin switching properties.
These are robust against disorder and edge roughness as present in experiments. 
Mediated through an effective spin orbit coupling arising in the constriction, a local
top gate enables switching between the edge states, while side gates allow for pure spin 
transistor action.
Such nano-constrictions may serve as building blocks and connectors for more
complex spin- and charge-selective edge channel networks based on topological insulators.

\section*{Acknowledgements}
This work is supported by Deutsche Forschungsgemeinschaft
(GRK 1570 and joined DFG-JST Forschergruppe Topological Electronics).
The authors thank H. Buhmann, M. Wimmer and J. Wurm for useful conversations
and M. Krueckl, J. Kuipers and M. Wimmer for a careful reading of the manuscript.

%\bibliography{paper}
%Merlin.mbs v4.21 2009-07-09.
%

\end{document}